\documentclass[10pt]{article} 

\usepackage[margin=3.5cm]{geometry}

\usepackage{amsthm,amsmath}
\usepackage{amsfonts}
\usepackage{makecell}

\usepackage{graphicx}

\usepackage{authblk}
\usepackage{url}

\newcommand{\csentence}[1]{#1}

\title{Balancing Capacity and Epidemic Spread in the Global Airline Network}
\date{}

\author[1,4]{Robert Harper}
\author[2,3,4]{Philip Tee}

\affil[1]{Science Group, Moogsoft Ltd., London, UK}
\affil[2]{Science Group, Moogsoft Inc., San Francisco, USA}
\affil[3]{The Beyond Center, University of Arizona, USA}
\affil[4]{Dept. of Informatics, University of Sussex, UK}

\begin{document}

\maketitle

\abstract{
The structure of complex networks has long been understood to play a role in transmission and spreading phenomena on a graph.
This behavior is difficult to model analytically and is most often modeled numerically.
Such networks form an important part of the structure of society, including transportation networks.
As society fights to control the COVID-19 pandemic, an important question is to choose the optimum balance between the full opening of transport networks and the control of epidemic spread.
In this paper we investigate how recent advances in analyzing network structure using information theory could inform decisions regarding the opening of such networks.
By virtue of the richness of data available we focus upon the worldwide airline network, but these methods are in principle applicable to any transport network.
We are able to demonstrate that it is possible to substantially open the airline network and have some degree of control on the spread of the virus.
}

\section{Introduction and Motivation}
\label{sec:introduction}

\subsection*{Background}
Since December 2019 the world has been adjusting to life with COVID-19 (officially SARS-CoV-2), with the first outbreak being reported in Wuhan in China \cite{wu2020outbreak}.
The nature of the disease is understood to be a highly contagious viral infection, causing severe respiratory complications and high fatality rates.
Societies worldwide are facing up to the first global pandemic since the `Spanish Flu' outbreak of 1918; a so-called once in a century event.
What is very evident from available data is that the epidemic will very sadly continue to claim lives until a cheap and simple treatment or vaccine is available.
At the time of writing deaths exceed one million individuals \cite{JHU2020}, and the `second wave' is well underway as we head into the northern hemisphere winter.
Public policy towards control of this disease has mostly focused upon social-distancing measures to break the chain of infection.
This capitalizes upon the idea that COVID-19 is spread via (at least approximately) person to person contact, and by reducing social mixing the epidemic will be significantly slowed.
This conclusion is based upon the well understood models of network epidemiology which in turn relies upon a contact graph \cite{pastor2001epidemic,newman2002spread,kiss2017mathematics} over which the network spreads.

Interruption of this network of social contacts via a lockdown has serious economic consequences.
In the initial phases of the pandemic the consequences of the lockdown included a drop $87.5\%$ in airline traffic in China accompanied with a drop of $21.2\%$ in retail sales \cite{Malden2020}, and a precipitous contraction of $20.4\%$ in the GDP of the United Kingdom \cite{Scruton2020} as an example.
Although economies recovered moderately in the summer of 2020, economic activity is still well below normal.

The purpose of this paper is to investigate how the two deleterious consequences of the pandemic may be balanced.
On the one hand it is unrealistic for economies to remain locked down, and on the other it is vital to control the speed of the epidemic to minimize fatalities.
At the heart of economies are transport networks, and the capacity of transport networks is an indicator of economic activity.
The question that we pose concerns how the reduction of capacity in a transport network effects the spread of an epidemic on the same network, as modelled by as a site percolation model described in Section~\ref{sec:methods}.
We focus upon the airline transport network as detailed data is available on routes \cite{oag,ourairports}, capacity and timetables, although the methods and analysis are applicable to any transport network.
It is interesting to note that certain \textit{Agent Based Models} (ABM) of epidemic spread focus upon airports as the principle point of ingress of a disease \cite{cliff2018investigating}, and so controlling the spread through the airport network would significantly reduce community spread inside of geographically isolated nations such as Australia.

Specifically, we are able to compare different schema for the reduction of capacity of the network on their effect of the rate of spread of the epidemic as measured by the effective reproduction rate of the epidemic, $R_e$.
We are careful to stress that this is an effective reproduction rate, not $R_0$, as the primary purpose of our model is to experiment with network restriction, not provide robust predictions of epidemic spread.
Nevertheless, we are able to show that there is a significant difference between random closure of airports and routes and a selective method that uses graph properties of the network to select airports and routes to close.

In analyzing the schema for route limitation in the transport network we focus upon the network structure.
It is well known that real world networks, particularly those possessing the scale-free property have non trivial behavior as links are progressively removed \cite{albert2000error}.
In particular, the random removal of links tends not to provoke functional failures of a network (as measured by the number of nodes that become unreachable), but targeted removal of hubs (nodes with a large degree or number of links) can provoke failure very quickly.
It is this asymmetric behavior under link suppression that provides the motivation for our approach, as the connectivity of a network has a profound impact on the speed with which an epidemic can spread, as modeled by a random walk of infectious individuals upon it.

In addition to investigating the removal of hubs of large node degree, we also explore the use of vertex entropy, an informational measure of node importance \cite{dehmer2008information,dehmer2011history,tee2017vertex,tee2018relating}, as a schema for selecting route reduction by closing airports.
Vertex Entropy is closely correlated with centrality measures, and does not in general focus on the hubs in a network, but instead the `pinch points' in connectivity.
We speculate that these nodes represent important pathways for spreading phenomena, but not necessarily high capacity routes in the network.

The relationship between $R_e$ and network capacity is complex and non-linear.
In the cases we examine, we show that for a drop in $R_e$ of $0.15$, the capacity of some of the underlying networks can be 50\% greater than for the random removal case.
We believe this justifies the principle of partial network restriction for epidemic moderation.

It is possible to refine further the methods used to model the spread of the disease in a much more granular fashion, but we believe these results form a motivation to produce these more detailed models and perhaps form the basis of an alternative approach to managing COVID-19 than repeated and complete closure of transport infrastructure.

\subsection*{Outline of this paper}

We begin in Section~\ref{sec:theory} with an overview of the necessary concepts of network epidemic models and vertex entropy in which to frame the experiments we undertake.
We describe the simulation and experimentation in Section~\ref{sec:methods}, outlining the construction of both our epidemic spreading model and also the route restriction methodology.
In Section~\ref{sec:results} we discuss the results obtained and we conclude in Section~\ref{sec:conclusions}, including an outlook for further work.

\section{Theoretical Considerations}
\label{sec:theory}

\subsection*{Network Epidemic Models}

As was noted in the introduction, infections require a physical means of transferring from an infectious individual to a susceptible one.
Some diseases, such as sexually transmitted ones, require physical contact, whereas airborne infections, believed to include COVID-19, only require proximity.
Whatever the biological mechanisms at the core of epidemic spread is a network of contacts.
A contact network represents individuals (or places) as nodes, and the links represent a contact along which transmission can occur.
The disease then proceeds by transmission along the links usually governed by a transmission probability.
The structure of the network has an important role on the progression of the network \cite{bell2020beyond,newman2002spread}, and provides the starting point for the strategy we investigate in this paper.

In particular, many real world networks are known to possess the `small world' property \cite{watts1998collective}, involving the presence of hubs that create short cuts in the network and dramatically reduce the graph diameter.
In principle, fewer network hops are required on average to reach a node, with obvious implications for epidemic spread.
This property can be replicated with networks that are generated by the various forms of preferential attachment \cite{barabasi2016network,Albert2002}, which produce a scale free degree distribution.
It is a well known claim that real world networks have a power law degree distribution whereby $P(k)\propto k^{-\alpha}$, with values of $\alpha$ typically in the range $2.0\text{\nobreakdash--}3.0$, $P(k)$ being the probability of a randomly chosen node having degree $k$.
This has been much disputed \cite{broido2019scale}, and indeed contact networks may not be scale free \cite{vanhems2013estimating}, but transport networks are hypothesized to exhibit reasonably strong scale freedom \cite{sridhar2008network}.
Our analysis of the data does not agree with this conclusion however.
Using the approach outlined in \cite{broido2019scale}, we obtain a weak scale-freedom with a power law exponent $\alpha=1.73$, and a better fit for a truncated power-law.
This result is intriguing, because although the airline network may not itself be scale-free, it does exhibit similar resilience behavior.

In our work we adapt the percolation approach to modeling epidemic spread \cite{moore2000epidemics,newman2002spread}.
The percolation model depends upon each link in a contact network permitting transmission of the disease according to a probability $T$ called the transmissibility.
Effectively, one starts with one infected node and then as a path of infection is traversed the link is marked as `occupied' and the component of the graph connected by such links emerges as a cluster representing the infection.
When this sub-graph captures a finite fraction of the nodes a \textit{Giant Component} (GC) emerges.

The transmissibility governs the size of the GC and is dependant upon the length of time an individual is infectious $\tau$, and the rate $\beta$ at which an individual infects one of its contacts.
Providing $\beta$ is independent on time, in terms of these parameters its value is
$T=1-e^{- \beta \tau}$.
As this probability varies from $0.0$ to $1.0$ the size of the GC does not vary smoothly, but in general transitions to a large fraction (say $> 50\%$) at a distinct value of $T$.
That is, the epidemic undergoes a phase transition \cite{dorogovtsev2008critical}, and this critical behavior is a function of the control parameter $T$.
Below a critical value $T_c$ the size of the GC is very small and the epidemic not widespread. 
Above $T_c$ the spread affects a finite fraction of the population.

We describe in Section \ref{sec:methods} how we apply transmissibilty and percolation to our simulation, but at coarse scale it is not meaningful, in a network carrying millions of passengers, to model every interaction between infectious and susceptible passengers.
Instead, we use the concept of a contact probability of transmission to determine the number of infected passengers that exit a flight dependant upon the capacity of the aircraft.
This is a key simplification (and vulnerability) of our model, and in future work we intend to expand our model to take account in a more granular fashion the person to person interactions of our infected individuals.

This detail notwithstanding, it is clear that the value of $\beta$ plays an important role whether the epidemic spreads at all on our toy model.
We assume for the purposes of exploring the effectiveness of our route closure selection schema a value of $\beta$ that will lead to an endemic spread of the infection.

\subsection*{Vertex measures of Graph Entropy}

The concept of graph entropy was first introduced by Janos K{\"o}rner in 1972 \cite{korner1986fredman,simonyi1995graph}.
Since then many approaches \cite{passerini2008neumann,bianconi2007entropy,bianconi2009entropy} have emerged to analyze and quantify the information encoded in the structure of a graph, in particular how the potentially vast configuration space of graphs that share common features (such as degree distribution) effectively `hide' information and therefore have entropy.
In essence, graph entropy measures the complexity of a graph and is neither easy nor efficient to compute.
For example, the original definition of K{\"o}rner relies upon the stable sets of a graph, a well known NP-complete problem.
For practical purposes it would be ideal if an approximate, vertex level measure of entropy were available.

One such approach, termed \textit{Vertex Entropy} (VE), was primarily defined on unweighted, simple graphs \cite{tee2017vertex}, based upon a formalism for vertex level entropy first introduced by Dehmer \cite{dehmer2008information,dehmer2011history}.
Dehmer utilized the concept of a \textit{local functional} for a vertex, which can be scoped to calculate values for every vertex based upon the local topology of the graph.
The degree of locality in the treatment is controlled by using the concept of the \textit{j\nobreakdash-sphere}, $S^{j}$, in the graph, centered at a given vertex.
A j\nobreakdash-sphere is a subset of vertices of distance $j$ from a given vertex $v_{i}$, where distance ${d(v_{i},v_{j})}$ is the fewest number of edges in a walk from $v_i$ to $v_j$.
This definition excluded the vertex $v_{i}$, and other interior nodes for ${j \geq 1}$, but this introduces problematic zeroes when we introduce the clustering coefficient.
Accordingly, in \cite{tee2017vertex}, we extended the definition to include the vertex $v_{i}$ as part of the set.
The definition of $S^{j}$ is then modified as follows.
For a graph $G(V,E)$, we define for a vertex ${v_{i} \in V}$, the j\nobreakdash-sphere centered on $v_{i}$ as $S^{j}_{i}= \{ v_{k} \in V \vert d(v_{i},v_{k}) \leq j, j \geq 1 \} \cup \{ v_{i} \}$.
The concept of j\nobreakdash-spheres is a convenient formalism to capture locality in the graph and by breaking a large graph into j\nobreakdash-spheres, we can progressively examine complex combinatorial quantities such as graph entropy on increasingly larger subsets of the graph.
We proceed by equipping each $S^{j}_{i}$ with a positive, real-valued function ${{f_{i} : v_{i} \in S^{j}_{i} \rightarrow \mathbb{R}^{+}}}$.
This function is proposed to be dependent upon properties of the nodes that are members of the j\nobreakdash-sphere, such as their degree, number of cycles and so on, which capture the local structural properties of the graph.
We can then construct a well defined probability function for each vertex,
\begin{equation}\label{eqn:functional}
	p_{i}=\frac{f_{i}} {\sum_{ v_{j} \in V } f_{j}} \text{.} \\
\end{equation}
which satisfies $\sum_{i} p_{i} = 1$.
These functions are then used to construct entropy measures in direct analogy to Shannon entropy, $H(v_{i}) = - p_{i} \log_{2} p_{i}$.
A basic form of vertex functional can be chosen to be $f_i=k_i$.
The probability in this instance, as defined by Eq.~\eqref{eqn:functional}, represents the probability of a randomly chosen edge being incident upon vertex $v_i$.
This choice gives the first definition of a VE,
\begin{equation}\label{eqn:frac_deg}
	H_k(v_{i})=\frac{k_{i} }{2 \lvert E \rvert } \log_{2} \left ( \frac{ 2 \lvert E \rvert }{k_{i} } \right ) \text{,}
\end{equation}
where $ \lvert E \rvert $ is the number of edges in the graph, recalling that ${\sum_i k_i = 2 \lvert E \rvert }$.
The definition of VE was analyzed in \cite{tee2017vertex} and termed \textit{Fractional Degree Entropy}.

The influence of a vertex within any graph is likely to be related to how it is connected into that graph, not simply by how many edges are incident upon it.
To capture this local structure we consider the concept of vertex clustering first introduced in \cite{watts1998collective} as a normalizing factor in our definition of a vertex entropy, such that $H^{\prime}(v_{i}) = \frac{ H(v_{i}) }{ C_{i}^{j} }$, where $C_{i}^{j}$ represents a local clustering coefficient for the j\nobreakdash-sphere of interest.
In this equation however, we modify the normal clustering coefficient \cite{watts1998collective} to include edges incident on $v_i$, and we define $C_{i}^{j}$ in terms of the number of edges in a j\nobreakdash-sphere edge set $\lvert E^{j}_{i} \rvert$ as,
\begin{equation}\label{eqn:local_cc}
	C_{i}^{j} = \frac{ 2 \lvert E_{i}^{j} \rvert }{ k_{i} ( k_{i} + 1 ) } \text{.}
\end{equation}
And so we have \textit{Normalized Fractional Degree Entropy} defined as,
\begin{equation}\label{eqn:norm_frac_deg}
	H_{k}^{\prime}(v_{i}) = \frac{ k_{i}^{2}( k_{ i } + 1 ) }{ 4 \lvert E_{i}^{j} \rvert \lvert E_{\vphantom{i}}^{\vphantom{j}} \rvert } \log_{2} \left ( \frac{ 2 \lvert E \rvert }{k_{i} } \right ) \text{,}
\end{equation}

\subsection*{Vertex Entropy for Weighted Graphs}

In the following section we introduce an extension to VE for a weighted graph $G(V,E)$ on $N =| V |$ vertices.
We define the weighted adjacency matrix $w_{ij}$, such that the weight of an edge, ${e_{ij} \in E}$, is defined as $w_{ij}$, where ${w_{ij} = 0}$ if there is no edge between $v_i$ and $v_j$, and $w_{ii}=0$.
As we are still dealing with undirected graphs, we can, instead of using the number of edges incident on a vertex, compute a weighted degree, $K_i$, such that
\begin{equation}\label{eqn:weighted_k}
K_{i} = \sum \limits_{ j = 0 }^{ j < N } w_{ij} \text{.}
\end{equation}

To produce a value probability we need the sum of this weighted degree.
As this sum will be proportional to $ 2 \lvert E \rvert $ we can write, $\sum \limits_i K_i = W \times 2 \lvert E \rvert$, where W is a constant.
It suffices now to normalize the weights, by dividing by $W$, and to define our vertex functional $p_{i}^{w}$, as follows, $p_{i}^{w} = \frac{ K_{ i } }{ W }$, with identically $\sum \limits_{i} p_{i}^{w} = 2 \lvert E \rvert$.
We now use the the well defined probability ${p_{i} = K_{i}/2 \lvert E \rvert }$ to define our \textit{Weighted Fractional Degree Entropy} as:
\begin{equation}\label{eqn:weight_frac_deg}
	H_K(v_{i})=\frac{K_{i} }{2 \lvert E \rvert } \log_{2} \left ( \frac{ 2 \lvert E \rvert }{K_{i} } \right ) \text{,}
\end{equation}
with its normalized counterpart given by:
\begin{equation}\label{eqn:norm_weight_frac_deg}
  H_{K}^{\prime}(v_{i})=\frac{ H_K(v_{i}) }{ C_{i}^{j} }\text{.}
\end{equation}

\section{Materials and Methods}
\label{sec:methods}

\subsection*{Data Source and Data Manipulation}

The airline route graph used in our analyses was constructed using flight data from \cite{oag} and airport data from \cite{ourairports}.
Each entry within the flight database includes departure and arrival airports, airline and flight number, flight duration, plane capacity, code\nobreakdash-share data, annual flight schedule etc.
Best efforts were made to remove duplicate entries arising from code-sharing and multi-leg flights.

We define a route as any pair of airports between which there is at least one flight.
For each route we aggregate flight details from across the year to give metrics such as flight distance, flights per day, passengers per day, and flight capacity.

More formally, we define our airline route graph as an undirected, weighted graph, $G( V, E, w_{ij} )$, where $V$ is the set of airports, $E$ the set of routes, and $w_{ij}$ the matrix of weights for individual routes such as its capacity in passengers.

\subsection*{Overall Objective}
Using this weighted graph we simulate the spread of a disease by modelling a random walk of infectious individuals across the route network.
As nodes are visited we mark the site as infected and continue our random walk, propagating the epidemic by using a coarse grained approach to infection using the concept of transmissibility.

The coarse graining is obtained by categorizing the routes by the average capacity of flights operating between the airports.
We assume that for each arriving flight containing one infectious individual the flight yields additional infectious individuals according to the capacity of the aircraft.
This of course is not a realistic assumption for a true model of epidemic spread, not least because the serial interval, being the time between becoming infected and infecting another person is estimated to be ${3\text{\nobreakdash--}7}$ days for COVID-19 \cite{park2020systematic}.
Our justification is that we are effectively `compressing' time by making this assumption, and we are principally interested in the overall effect of network changes on the spreading phenomenon, not detailed predictions for the actual spread of the disease~--- that would require a much more detailed model such as an ABM \cite{cliff2018investigating}.
We compute over averages of the runs it takes for $80\%$ of the nodes to become affected and then historically compute the $R_e$ value of the infection.
The computation of $R_e$ is undertaken by optimizing a fit for the early part of the infection using an exponential spread model and extracting the reproduction rate.

\subsection*{Simulation Detail}

Every random walk starts at a random node in the graph.
The onward step at every stage of every walk is chosen using a probability weighted by the number of passengers travelling along each route from the current location.
Intuitively, this describes the case of a passenger randomly picking a single ticket from the set of all tickets available across all flights for all destinations available from that airport.

Our analyses consider five approaches to restricting the size of the network.
The first regime is based upon random selection.
A further four use targeted approaches based upon the local structure of the graph: node degree; VE, Eq.~\eqref{eqn:norm_frac_deg}; weighted degree, Eq.~\eqref{eqn:weighted_k}; and weighted VE, Eq.~\eqref{eqn:norm_weight_frac_deg}.
We use the number of passengers along each route as the weighting criteria.
For random removal, and in order to give a more realistic simulation, we consider only those airports categorized as \textit{large} by the `ourairports' data \cite{ourairports}.

The number of onward infections is chosen in a two-stage process.
Our simulation requires discrete walkers.
To facilitate this we map a nominal value for $R_e$ (in this case $1.5$), onto a discrete probability density function and randomly choose values from that distribution.
In order to introduce more realism into the transmission model we factor this value based on the average capacity of the planes serving the route.
We use a factor of $2$ for medium aircraft ($150\text{\nobreakdash--}300$ passengers) and a factor of $3$ for larger aircraft.
We acknowledge the limitations of this approach.

To cover the different node removal regimes we adopt two subtly different approaches to the random walks.
The common elements of both are: nodes are removed from the graph until its size reduces by $20\%$ from its original size; and each random walk terminates once $80\%$ of the graph has become infected.
For the targeted approaches we start by sorting all nodes in descending order of the value of the removal criteria i.e. degree, VE, weighted degree and weighted VE.
For each node that is removed we conduct $100$ random walks.
The random removal regime is more complex as there are two sources of randomness: the random walk itself, and the choice of nodes used to restrict the network.
We restrict the graph by randomly removing nodes, for each node that is removed we execute $50$ random walks upon the resulting graph.
This process is repeated $20$ times.\\

In practice, we do this by `closing' airports and all associated routes.
As we do this we can compute the passenger carrying capacity of the remaining network as measured by passenger\nobreakdash-kms.
Our experiment is to investigate the relationship between capacity and $R_e$ using these different schema, with the objective of identifying an optimum method of network restriction.

\section{Simulations and Results}
\label{sec:results}

In the following section we present the results of our simulations and highlight the key observations.
Our primary objective is to study the impact of five different node removal regimes on epidemic propagation, specifically: random removal of nodes representing large airports and targeted removal based on the weighted and unweighted forms of degree and vertex entropy.
All other free-parameters and simulation techniques such as the transmission mechanism, random walk stopping criteria, route selection etc. remained the same as described in Section~\ref{sec:methods}.

\begin{figure}[hbt]
  \centering
  \includegraphics[width=0.6\textwidth]{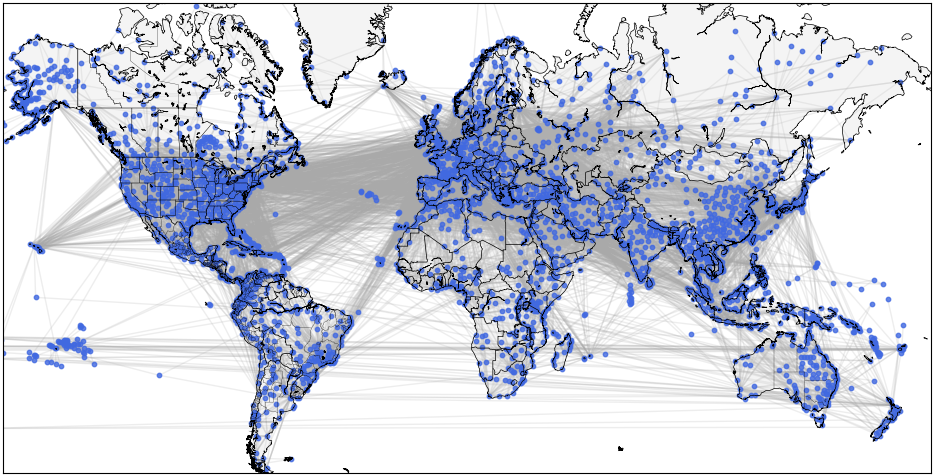}\\  
  \includegraphics[width=0.45\textwidth]{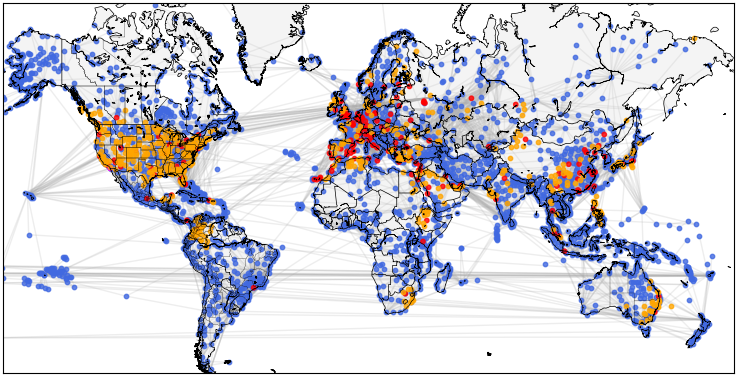}  
  \includegraphics[width=0.45\textwidth]{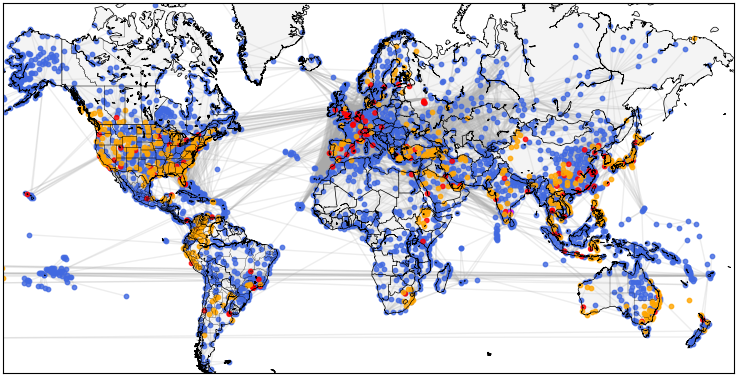}  
  \caption[]{\csentence{Global Airline Route maps under different node removal strategies.} a) The complete route graph. b) Nodes removed by degree. c) Nodes removed by Vertex Entropy}
  \label{fig:maps}
\end{figure}

Fig.~\ref{fig:maps} shows the qualitative impact of two different node removal strategies.
Fig.~\ref{fig:maps}a shows the map of all airports and routes in the global network prior to any node removal.
Each dot represents an airport and each line a route.
Figs.~\ref{fig:maps}b and \ref{fig:maps}c show the impact of reducing the size of the original network by $20\%$ using degree and vertex entropy respectively.
Red dots represent airports that have been removed from the network, while yellow dots indicate airports that have subsequently become disconnected from the core of the graph.

In both removal scenarios the impact on global routes is profound with large reductions in open routes across the world.
Most clear is the reduction in routes across Southern-Asia and in the trans-Atlantic regions.
Looking more closely at Figs.~\ref{fig:maps}b and \ref{fig:maps}c, it is interesting to note that the different removal regimes have impacted different geographical regions.
The quantity of removed and disconnected airports is noticeably larger for degree-based removal in North America and Europe.
On the other hand, VE-based removal has a more uniform impact, resulting in more nodes being removed or disconnected in South America and South-East Asia.

The quantitative impact of the different strategies is shown in Table~\ref{tab:node_removal_metrics}.
Comparison of the unweighted degree and VE-based removal strategies show that ${\sim}36\%$ more nodes need to be removed in the degree-based case to achieve a $20\%$ reduction in the size of the GC.
This is consistent with the objective of VE as a measure to identify single points of failure more accurately than degree.
In both cases the number of edges falls, very roughly, by two\nobreakdash-thirds, however, even more striking is the corresponding ${\sim}90\%$ collapse in network capacity.
For the VE regime the number of edges and the network capacity are both approximately $25\%$ higher than for the degree-based case.

\begin{table}[hbt]
  \centering
  \caption{Network metrics under different node removal strategies.}
  \label{tab:node_removal_metrics}
  \begin{tabular}{cccccc}
    \hline
    Node Removal Strategy & \thead{Removed\\Nodes} & \thead{Disconnected\\Nodes} & \thead{Retained\\Nodes} & \thead{Retained\\Edges} & \thead{Retained Capacity\\$\times10^9$ pkm} \\
    \hline
    None & 0 & 0 & 3661 & 24683 & 9.74 \\
    Degree & 191 & 540 & 2930 & 7805 & 0.80 \\
    Vertex Entropy & 140 & 592 & 2929 & 9774 & 1.06 \\
    Weighted Degree & 165 & 567 & 2929 & 10011 & 0.65 \\
    Weighted Vertex Entropy & 154 & 576 & 2931 & 9938 & 0.63 \\
    \hline
  \end{tabular}
\end{table}

The results for the weighted variants of degree and VE are closer to each other than for the unweighted forms.
The VE variant requires ${\sim}7\%$ fewer nodes to be removed, however, the number of edges in the resulting graphs differ by less than ${\sim}1\%$.
This would suggest a high degree of similarity between the sets of removed nodes in the weighted cases, we revisit this observation below.
In regard to capacity, the overall reduction is slightly higher at ${\sim}94\%$.
Given that the weighted removal strategies use metrics based upon passenger counts, this higher reduction in capacity may be expected.
What is perhaps surprising is that the unweighted measures, that rely only upon network structure, produce such similar reductions.

When we compare the results for the weighted and unweighted strategies with each other, different observations emerge.
More routes remain in the weighted cases but with a lower overall capacity.
This suggests that the weighted approaches tend to retain more airports, but airports that serve shorter, less busy routes.

\begin{figure}[hbt]
  \centering
  \includegraphics[width=.96\textwidth, height=.42\textwidth]{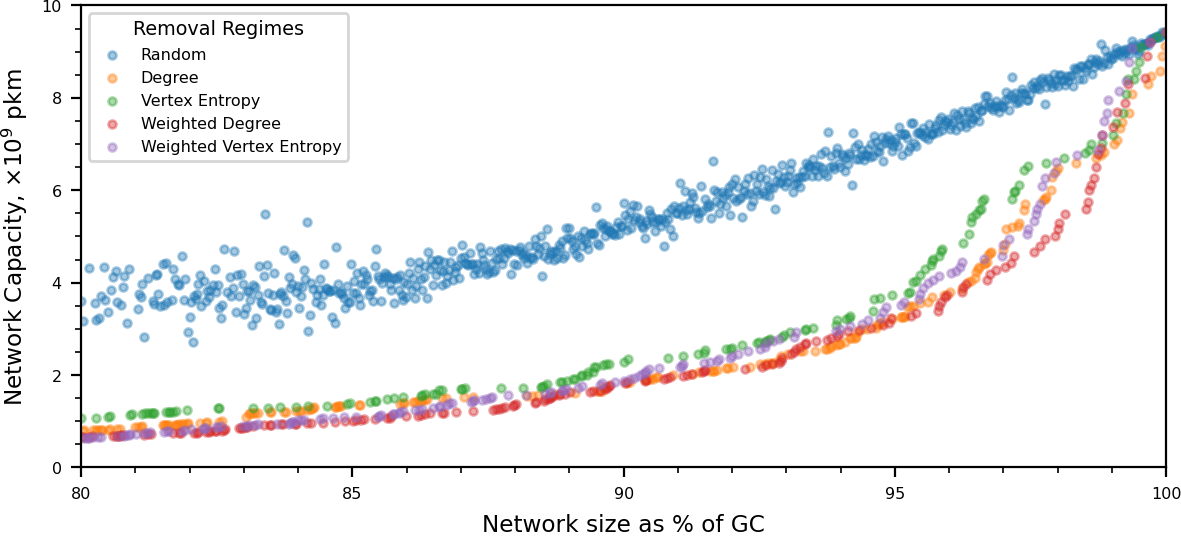}
  \caption[]{\csentence{Network capacity vs network size, under different node removal strategies.}}
  \label{fig:cap_by_nodes}
\end{figure}

The impact of different strategies on network capacity as nodes are removed can be seen in Fig.~\ref{fig:cap_by_nodes}.
For the case of random removal, capacity reduces approximately linearly as the network reduces in size down to sizes of ${\sim}85\%$.
As the size of the network reduces further the rate of reduction in capacity flattens.
The degree of scatter in the data also increases at this point, owing to the smaller sample sizes inherent to the random removal process and as we approach the $80\%$ threshold.

All of the targeted removal strategies exhibit similar characteristics.
The very steep reduction in capacity for only small reductions in network size is particularly apparent.
In fact a reduction in the size of the network of only a few percent can reduce its capacity by half.
Below GC sizes of ${\sim}95\%$, the rate at which capacity drops is reduced but in all cases the capacity at the $80\%$ cutoff lie in the range ${0.6\text{--1.0}\times10^9 \text{pkm}}$.
A striking feature of Fig.~\ref{fig:cap_by_nodes} is the small plateau in capacity for VE near $97.5\%$.
This could be because vertex entropy is disconnecting leaf nodes in the route graph by removing the local hubs that they connect into, which will have high vertex entropy but potentially carry few passengers.

\begin{figure}[hbt]
  \centering
  \includegraphics[width=.96\textwidth, height=.42\textwidth]{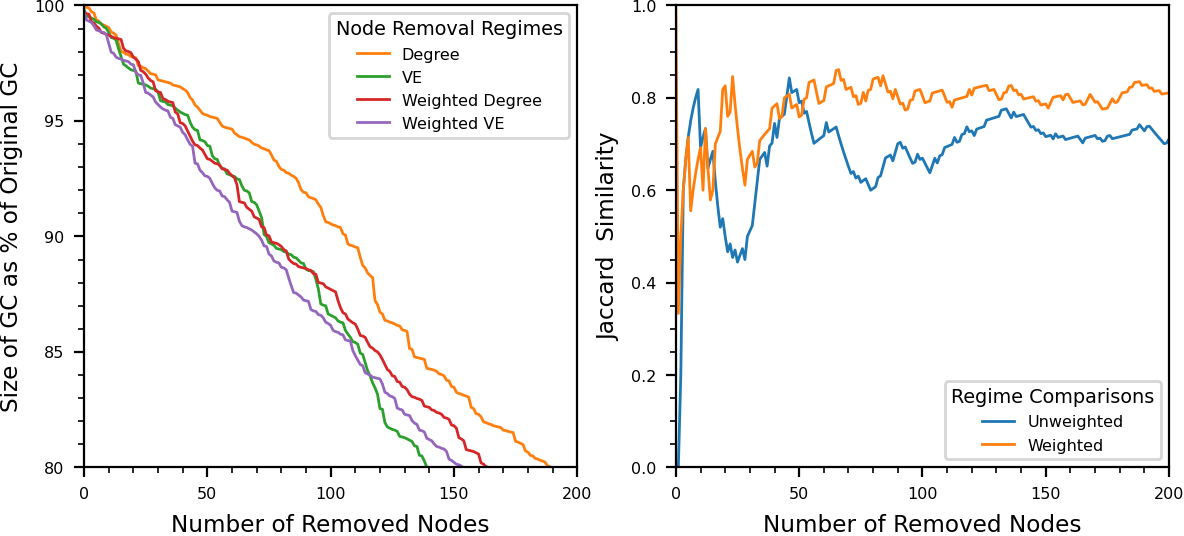}  
  \caption[]{\csentence{Comparison of the removed node-sets for different node removal strategies.}}
  \label{fig:rm_node_comps}
\end{figure}

We investigate the similarity of the removed node sets by examining how the size of the GC changes as we remove nodes in Fig.~\ref{fig:rm_node_comps}a and compare the specific node sets identified by the unweighted and weighted removal regimes in Fig.~\ref{fig:rm_node_comps}b using the well-known \textit{Jaccard similarity coefficient} (JC). 

Fig.~\ref{fig:rm_node_comps}a shows that the GC collapses at about the same rate for both weighted strategies.
More interesting is the variation in JC.
There is a clear variation in the node sets for about the first $50$ nodes to be removed but apart from the small peak at about $25$ nodes the JC exhibits a consistent upward trend to a limit of about 0.8.
We can conclude that a relatively large proportion of the nodes identified by the weighted regimes are in fact the same.

The unweighted cases behave somewhat differently.
Fig.~\ref{fig:rm_node_comps}a shows that the number of nodes required for the VE-based regime is consistently and considerably less than for the degree-based regime and as such the GC collapses at a much faster rate.
The behavior of JC, Fig.~\ref{fig:rm_node_comps}b, is quite erratic and far less consistent than for the weighted regimes.
There are multiple peaks and troughs across the full range of removed nodes suggesting that the unweighted techniques identify a more disparate set of nodes.
Interestingly, the trough in JC at about $25$ nodes corresponds to the plateau in network capacity observed in Fig.~\ref{fig:cap_by_nodes} at a GC size of $97.5\%$.

\begin{figure}[hbt]
    \centering
    \includegraphics[width=.96\textwidth]{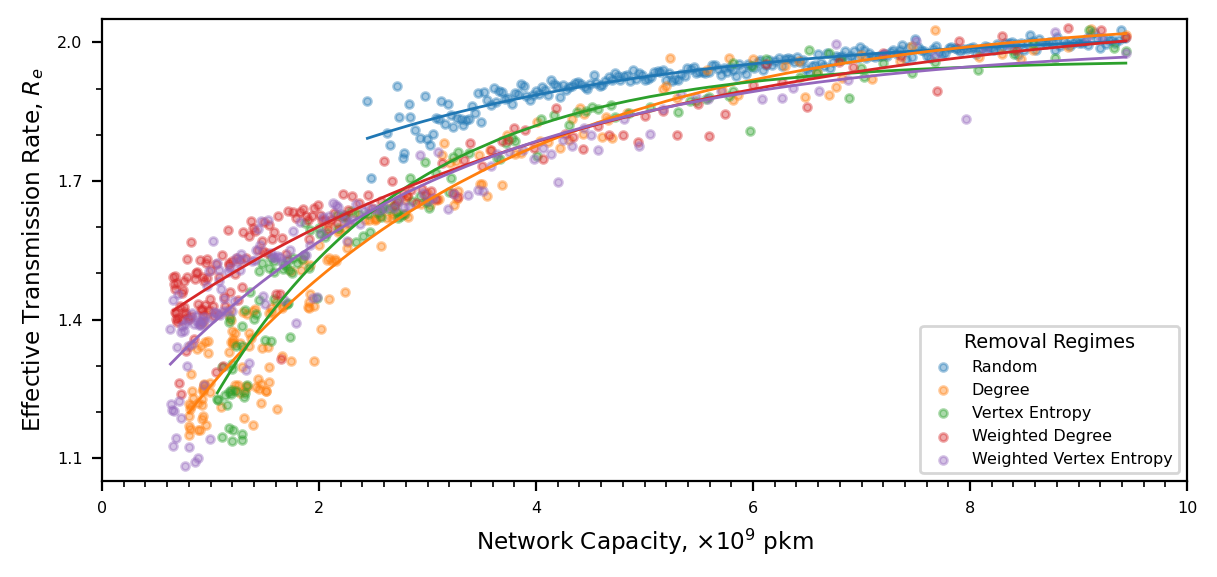}  
    \caption{\csentence{The impact of different node removal strategies on the relationship between network capacity and effective transmission rate.}}
    \label{fig:r_cap_with_fit}
\end{figure}

Fig.~\ref{fig:r_cap_with_fit} shows how the transmission rate on the network varies with its passenger carrying capability.
In all removal scenarios, a reduction in the capacity of the network reduces the transmission rate.
Unsurprisingly, random node removal has the least impact.
Indeed, even when reducing network capacity by two-thirds the reduction in effective reproduction rate $R_e$ is only $7.5\%$ to $1.85$.
At the same network capacity, all of the targeted node removal regimes show a far more substantial drop in transmission rate to about $1.72$, an overall reduction of about $14\%$ and almost double the reduction achieved by random removal.

An alternative analysis is to examine required capacity to achieve a target transmission rate.
Inspecting Fig.~\ref{fig:r_cap_with_fit}, for random node removal, a reduction of $0.15$ in $R_e$ can be achieved by reducing network capacity to ${3.2\times10^9}$ pkm.
For the same reduction in $R_e$, targeted node removal requires a capacity of about $5.0\times10^9$ pkm, a significant improvement.
We conclude that we can achieve the same $R_e$ but retain about $50\%$ more capacity using targeted node removal.
Cross-referencing these capacities with Fig.~\ref{fig:cap_by_nodes} shows that about $650$ nodes need to be removed or disconnected using random-removal.
Under targeted removal, and depending upon the specific regime, only about $70\text{\nobreakdash--}140$ nodes need to be removed.

Further examination of Fig.~\ref{fig:r_cap_with_fit} shows that for capacities above about $3.0\times10^9$ the targeted removal regimes behave similarly.
Below $3.0\times10^9$, their behaviors diverge a little, the unweighted forms giving larger reductions in transmission rate.
This behavior is consistent with our earlier observation regarding weighted regimes retaining more flights and hence providing more opportunity for transmission to occur.
We also observe that at the lowest capacities, both VE-based regimes produce larger reductions in transmission than their degree-based counterparts.

\section{Conclusions}
\label{sec:conclusions}

In our analysis we have demonstrated that it is possible to choose a scheme for route restriction in transport networks that optimizes capacity whilst restricting epidemic spread.
In particular, by using either a weighted vertex entropy or degree scoring of airports we are able to reduce the effective reproduction rate of an epidemic spread whilst preserving capacity much more effectively than with a random airport and route closure approach.
Owing to simplifications in the model, we should not place too much reliance upon the absolute values of $R_e$.
In relative terms, and for a reduction in $R_e$ of $0.15$, targeted node removal yielded networks with $50\%$ more capacity than those restricted by random selection.

This result would be perhaps less surprising if the airline network were robustly scale-free, but as stated before this is not conclusively true.
For scale free networks it is well understood that the `friendship paradox' \cite{feld1991your} exploits high degree hubs to propagate spreading phenomena.
Further, although vertex entropy is defined as a function of degree, the entropy based route restriction has a very different effect upon the capacity and epidemic in the low restriction regime.
Our result perhaps exposes a subtle interplay between network function and structure.

The purpose of our work is to motivate a discussion on how we can perhaps moderate our approach to extreme shutdown measures as a way to manage the public health challenges of COVID-19.
We believe there are many ways to improve the robustness of our model, and in further work we intend to refine our results by taking a route by route approach to restrictions and also implementing a finer grained spreading model.
In particular, we have not constrained the route restriction by other criteria such as retaining, or conversely severing, communications between particular geographic regions.
These additional constraints would likely change radically the conclusions of our experiments.
Despite the acknowledged shortcomings, we feel that our result justifies further exploration.

\bibliographystyle{plainurl} 

\bibliography{AirlineEpidemic}

\end{document}